\documentclass[journal,onecolumn,12pt,draftclsnofoot]{IEEEtran}
\makeatletter
\def\ps@headings{%
\def\@oddhead{\mbox{}\scriptsize\rightmark \hfil \thepage}%
\def\@evenhead{\scriptsize\thepage \hfil \leftmark\mbox{}}%
\def\@oddfoot{}%
\def\@evenfoot{}}
\makeatother
\ifCLASSINFOpdf
\else
\fi
\usepackage{cite}      % Written by Donald Arseneau
 
\usepackage{amsmath}
\usepackage{amssymb}
\usepackage{epsf}
\usepackage{color}
\usepackage{graphicx}  % Written by David Carlisle and Sebastian Rahtz
 
\usepackage{subfigure} % Written by Steven Douglas Cochran
\usepackage{url}
\usepackage{fancyhdr}
\usepackage[latin1]{inputenc}

\begin{document}
\title{Collaboration and Coordination in Secondary Networks for Opportunistic Spectrum Access}

%\author{
%\IEEEauthorblockN{Wassim Jouini}
%\IEEEauthorblockA{SUPELEC, SCEE/IETR\\
%Avenue de la Boulaie, CS 47601, \\
%35576 Cesson Sevigne Cedex, France\\
%Email: wassim.jouini@supelec.fr}
%\and
%\IEEEauthorblockN{Marco Di Felice, Luciano Bononi}
%\IEEEauthorblockA{Department of Computer Science,\\
%University of Bologna, Italy\\
%Email: \{difelice,bononi\}@cs.unibo.it
%}
%\and
%\IEEEauthorblockN{Christophe Moy}
%\IEEEauthorblockA{SUPELEC, SCEE/IETR\\
%Avenue de la Boulaie, CS 47601, \\
%35576 Cesson Sevigne Cedex, France\\
%Email: christophe.moy@supelec.fr}
%}

%\author{Michael~Shell,~\IEEEmembership{Member,~IEEE,}
%        John~Doe,~\IEEEmembership{Fellow,~OSA,}
%        and~Jane~Doe,~\IEEEmembership{Life~Fellow,~IEEE}% <-this % stops a space
%\thanks{M. Shell is with the Department
%of Electrical and Computer Engineering, Georgia Institute of Technology, Atlanta,
%\thanks{Manuscript received April 19, 2005; revised January 11, 2007.}}

\author{Wassim Jouini,
       Marco Di Felice,
      Luciano Bononi,
       and~Christophe~Moy% <-this % stops a space
\thanks{Wassim Jouini and Cristophe Moy are with the SUPELEC, SCEE/IETR, Avenue de la Boulaie, 35576 Cesson Sevigne Cedex, France, emails: \{wassim.jouini, christophe.moy\}@supelec.fr}% <-this % stops a space
\thanks{Marco Di Felice and Luciano Bononi are with the Department of Computer Science, University of Bologna, emails: \{difelice, bononi\}@cs.unibo.it}}
%\thanks{Manuscript received April 19, 2005; revised January 11, 2007.}}



\maketitle

\begin{abstract}

In this paper, we address the general case of a coordinated secondary network willing to exploit communication opportunities left
vacant by a licensed primary network. Since secondary users (SU) usually have no prior knowledge on the environment, they need to learn the availability of each channel through sensing techniques, which however can be prone to detection errors. We argue that cooperation among secondary users can enable efficient learning and coordination mechanisms in order to maximize the spectrum exploitation by SUs, while minimizing the impact on the primary network.  To this goal, we provide three novel contributions in this paper. First, we formulate the spectrum selection in secondary networks as an instance of the Multi-Armed Bandit  (MAB) problem, and we extend the analysis to the collaboration learning case, in which each SU learns the spectrum occupation, and shares this information with other SUs. We show that collaboration among SUs can mitigate the impact of sensing errors on system performance, and improve the converge of the learning process to the optimal solution. Second, we integrate the learning algorithms with two collaboration techniques based on modified versions of the  Hungarian algorithm and of the Round Robin algorithm, that allows to greatly reduce the interference among SUs.  Third, we derive fundamental limits to the performance of cooperative learning algorithms based on Upper Confidence Bound (UCB) policies in a symmetric scenario where all SU have the same perception of the quality of the resources. Extensive simulation results confirm the effectiveness of our joint learning-collaboration algorithm in protecting the operations of Primary Users (PUs), while maximizing the performance of SUs. 

\end{abstract}
% 
%\begin{IEEEkeywords}
%Coordination, Collaboration, Multi-Armed Bandit, Reinforcement learning, Multi-users, Multi-gamblers, Cognitive Radio, Opportunistic Spectrum Access, Upper Confidence Bound Algorithm, Imperfect Sensing, Sensing Errors.
%\end{IEEEkeywords}



% \fi
\IEEEpeerreviewmaketitle

%-------------------------------------INTRODUCTION------------------------------------------------ 
\section{Introduction}
%-------------------------------------------------------------------------------------------------
\newtheorem{theorem}{Theorem}
\newtheorem{lemme}{Lemma}
\newtheorem{corollary}{Corollary}
\newtheorem{definition}{Definition}
\newtheorem{scenario}{Scenario}
\newtheorem{fact}{Fact}
\newtheorem{policy}{Channel Selection Policy}
\newtheorem{coordination}{Coordination}
%\thispagestyle{plain}
%-------------------------------------------------------------------------------------------------
%\subsection{Opportunistic Spectrum Access}
%-------------------------------------------------------------------------------------------------

\par The general concept of Opportunistic Spectrum Access (OSA) defines two types of users: primary users (PUs) and secondary users (SUs). PUs access spectrum resources dedicated to the services provided to them, while SUs refer to a pool of users willing to exploit the spectrum resources unoccupied by PUs,  at a particular time in a particular geographical area, referred to as \textit{communication opportunities} \cite{FCC2002, Haykin2005}. 
\par The detection of opportunities and their exploitation in secondary networks can be challenging.  On the one hand, the secondary users can have different perceptions of a same opportunity depending on their observation abilities. Thus, a channel available with high probability -offering substantial communication opportunities- could be discarded by a SU unable to properly detect PUs' activity. On the other hand, several SUs can be competing for the same resources. Consequently, high interference can occur among them degrading the observed quality of the resources and the realized performance of the secondary network.

\vspace{0.3cm}

\par This study addresses the spectrum allocation problem in secondary networks, through the key concepts of \textit{learning}, \textit{collaboration} and \textit{coordination}. In order to implement the OSA paradigm in an efficient way, the SUs must be able to detect the
 the communications opportunities left vacant by incumbent users. Since usually no prior knowledge is available on the occupancy pattern of the channels, \textit{learning} abilities are needed. Several machine learning-based techniques have been proposed for spectrum allocation in secondary networks. Among these, Multi-Armed Bandit (MAB) techniques~\cite{Auer2002} have gained increasing interest, due to the possibility to derive  theoretical bounds on the performance of optimal learning algorithms. However, the impact of individual sensing error on the convergence of the learning algorithm is far to be completely explored. For this reason, in this paper we consider a collaborative network environment, where the  secondary users can collaborate and share the information learnt on the occupancy pattern of the channels. \textit{Collaboration} is a key element in Cognitive Radio (CR) networks \cite{CRN_2007,FuturD_2009}. Here, we investigate if and how the utilization of collaborative techniques can enhance the performance of the learning schemes, in order to enable secondary users to fully and quickly exploit vacant resources. At the same time, while collaborative learning is fundamental to mitigate the impact of PU interference, \textit{coordination} among SUs is required to guarantee optimal sharing of spectrum resources and to mitigate SU interference. The coordinator entity can be either real or virtual, but it should guarantee that -in the optimal configuration - a single SU is allocated per-channel.

\par In this paper, we introduce and analyze a joint coordination-learning mechanism. We state that the suggested mechanism enables secondary networks to deal with dynamic and uncertain environment in spite of sensing errors. We propose three novel contributions in this paper. First, we formulate the spectrum allocation problem in secondary networks as a special instance of the Multi-Armed Bandit (MAB) problem and we propose to solve it through algorithms derived by Upper Confidence Bound (UCB) policies~\cite{Poor2008,Jouini2009,Jouini2010}.  Compared to previous applications of MAB techniques on OSA issues, we address the case of  cooperative learning,  i.e.  SUs share the rewards in order to speedup the convergence of the learning algorithm to the optimal solution. Second, while learning PUs' occupation patterns of each spectrum band, we consider two general coordination algorithms whose purpose is to allocate at every iteration a unique SU per channel, in order to nullify the interference among SUs.  The coordination algorithm rely on a modified Hungarian algorithm \cite{Kuhn1955} and Round Robin algorithm, respectively, and our  modifications aim at providing a fair allocation of the resources. Third, we derive some fundamental results on the performance of collaborative learning schemes  for spectrum selection in secondary networks. More specifically, we demonstrate that -in a symmetric scenario where all SU have the same perception of the quality of the resources (yet with sensing errors)~\footnote{We refer to this scenario as \textit{symmetric} or \textit{homogeneous} scenario in the following.}   the $UCB_1$ algorithm can efficiently learn accessing optimal solutions even without prior knowledge on the sensors performance. Both results, in the case of symmetric and non-symmetric environments are illustrated through extensive simulations. 

\vspace{0.3cm}
\par The rest of this paper is organized as follows.
\par Section~\ref{related} discusses the works related to this paper and found in the open literature. Section \ref{sec:model_MOSA} details the considered OSA framework in this paper. To deal with uncertainty, a collaborative learning mechanism is proposed in Section \ref{sec:lm}.
The considered coordination mechanisms are modeled as instances of  Job Assignment problems, and are detailed in Section \ref{sec:geap}. The theoretical analysis of the joint learning-coordination framework is discussed in  Section \ref{sec:theory}.  Section \ref{sec:info} describes the collaboration mechanisms implicated in this OSA context. Finally, Section \ref{sec:simu} empirically evaluates the introduced coordination and learning mechanisms, while Section \ref{sec:concl} concludes the paper.

%-------------------------------------------------------------------------------------------------
\section{Related work}
\label{related}
%-------------------------------------------------------------------------------------------------
\par Several authors have already proposed to borrow algorithms from the machine learning community to design strategies for SUs that can successfully exploit available resources. We focus this brief overview on MAB related models applied to OSA problems.
\par To the best of our knowledge, the first extensive work that tackles spectrum band allocation under uncertainty applied to OSA, was presented in  \cite{Poor2008}. The paper presented various models where a single or multiple secondary user(s) aim(s) at opportunistically exploiting available frequency bands. Among other models, a MAB model was suggested in the case of perfect sensing (i.e., the state of a sensed channel is acquired without errors). The authors of \cite{Poor2008}  suggested the use of the algorithm $UCB_1$ and extended its results to the case of multi-channel selection by a single user. The case of multi-secondary users was also discussed. However a game theory based approach was suggested to tackle the problem. Such approaches lead to asymptotic Nash equilibrium, that is known to be difficult to compute in practice.
\par Since then, several papers suggested MAB modeling to tackle OSA related problems. In \cite{Jouini2009,Jouini2010}, the authors compared $UCB_1$ and $UCB_V$ algorithm \cite{Munos2007,Audibert2009} in the context of OSA problems, while \cite{LiuImperfectSensing2010,Liu2010} suggested to tackle multi-secondary users OSA problems modeled within a MAB framework. The algorithm analyzed in  \cite{LiuImperfectSensing2010,Liu2010} was borrowed from \cite{Robbins1985}. This algorithm is designed for observations drawn from Bernoulli distributions and known to be asymptotically optimal in the case of one single user. Thus, to adapt to OSA contexts, they extended the results to multi-users first. Then proved that mild modification of the algorithm, that take into account the frequencies of the errors (i.e., false alarms and miss detection), maintain the order optimality of their approach. Finally, they also considered the case of decentralized secondary networks and proved their convergence asymptotically. 
\par Taking the sensing errors into account is a fundamental step to achieving realistic OSA models. However considering that the error frequencies are perfectly known can be limiting in some scenarios \cite{Fishman1992,Tandra_JSTSP_2008,JouiniSensing2011}. In  \cite{Jouini2011,Jouinibis2011}, the authors showed that $UCB_1$ does not require prior knowledge on the sensors' performance to converge. However, the authors showed that the loss of performance is twofold. On the one hand, false alarm (i.e., detection of a signal while the band is free) leads to missing communication opportunities. On the other hand, they also lead to slower convergence rates to the optimal channel. Relying on these results, the authors of \cite{Felice2011} provided complex empirical evaluations to estimate the benefit of $UCB_1$ combined with various multi-user learning and coordination mechanisms (such as softmax-UCB approach for instance). 
\par Within a similar context, an interesting contribution can be found in \cite{Anandkumar2010}. They analyzed, in the case of errorless sensing, the performance of $UCB_1$ algorithms in the context of several secondary users competing to access the primary channel. No explicit communication or collaboration is considered in this scenario, yet, once again, $UCB$ algorithms are proven to be efficient to handle this scenario and to have an order optimal behavior.
\par All hereabove mentioned paper, consider homogeneous environment (or sensing). Namely, the frequency errors for all users and through all channels are the same. An exception can be found in  \cite{Gai2010,Gai2010bis}. As a matter of fact, they provided a general heterogeneous framework. It is worth mentioning that these papers do not consider a specific OSA framework. They rather consider that the observed expected \textit{quality} of a resource can be different. Consequently, the suggested model tackles multi-users in a general MAB framework rather than a specific application. The model, referred to as combinatorial MAB framework, is solved relying on a modified version of $UCB_1$ algorithms and the Hungarian algorithm. %Eventually, Papers \cite{Gai2010,Gai2010bis} showed that $UCB_1$ algorithms can efficiently adapt and provide efficient learning for MAB resource allocation problems. 
\par The work ~\cite{Gai2010bis} is the closest to the one provided within this paper. Unfortunately, since their model presents a general framework, it does not explicitly take into account the impact of sensing errors, nor does it show how would perform the algorithm in the case of collaborative homogeneous networks. Moreover, the Hungarian algorithm was only introduced as a possible optimization tool to solve their mathematical problem, but it was not considered form a network coordination perspective. The latter aspect is addressed by this paper.
\section{Network model}
\label{sec:model_MOSA}
%-------------------------------------------------------------------------------------------------

\par In this section we detail the considered OSA framework.
%---------------------------
\subsection{Primary Network}
%---------------------------
\label{subsec:PN}
\par The spectrum of interest is licensed to a primary network providing $N$ independent channels. We denote by $n \in \mathcal{D}=\{1,\cdots, N\}$ the $n^{th}$ channel. Every channel $n$ can appear, when observed, in one of these two possible states \{idle, busy\}. In the rest of the paper, we associate the numerical value $0$ to a busy channel and $1$ to an idle channel.  The temporal occupancy pattern of every channel $n \in \mathcal{D}$ is thus supposed to follow an unknown Bernoulli distribution $\theta_n$. Moreover, the distributions  $\Theta=\{\theta_1, \theta_2, \cdots, \theta_N\}$ are assumed to be stationary.  
\par In this paper we tackle the particular case where PUs are assumed to be synchronous and the time $t=0,1,2\cdots,$ is divided into slots. We denote by $\mathbf{S}_t$ the channels' state at the slot number $t$:  $\mathbf{S}_t=\{S_{1,t}, \cdots, S_{N,t}\} \in \{0,1\}^N$. For all $t\in\mathbb{N}$, the numerical value $S_{n,t}$ is assumed to be an independent random realization of the stationary distributions $\theta_n \in {\Theta}$. Moreover, the realizations $\{S_{n,t}\}_{t\in\mathbb{N}}$ drawn from a given distribution $\theta_n$ are assumed to be independent and identically distributed. The expected availability of a channel is characterized by its probability of being idle. Thus, we define the availability $\mu_n$ of a channel $n$, for all $t$ as: 
$$\mu_n {\buildrel \Delta \over =} \mathbb{E}_{\theta_{n}}\left[S_{n,t}\right] = \mathbb{P}\left(\text{channel $n$ is free}\right)=\mathbb{P}\left(S_{n,t}=1\right)$$
 
%------------------------------------------
\subsection{Secondary Users model}
%------------------------------------------
\label{subsec:SU}

We detail in this subsection the generic characteristics of all considered SUs.
\par  We consider $K$ SUs denoted by the index $k\in \mathcal{K}=\{1, \cdots, K\}$. At every slot number $t$, the SU has to choose a channel to sense.  To do so, the SU relies on the outcome of past trials. We denote by $i^{(k)}_t$ the gathered information until the slot $t$ by the $k^{th}$ SU. We assume that all SUs can only sense and access one channel per slot. Thus selecting a channel by a SU $k$ can be seen as an action $a^{(k)}_t \in \mathcal{A}$ where the set of possible actions $\mathcal{A}\subseteq\mathcal{D}=\{1,2,\dots, N\}$ refers to the set of channels available. In this paper, all SUs collaborate through a coordination mechanism described in  Section~\ref{sec:geap}. This latter, through either a centralized or decentralized approach allocates at every iteration $t$ a different channel to each SU.

\par The outcome of the detection phase is denoted by the binary random variable $X^{(k)}_t \in \{0,1\}$, where $X^{(k)}_t=0$ denotes the detection of a signal by the $k^{th}$ SU and $X^{(k)}_t=1$ the absence of a signal, respectively. In the case of perfect sensing, \mbox{$X^{(k)}_t=S_{a^{(k)}_{t},t}$} for all SUs, where $a^{(k)}_{t}$ refers to the channel selected at the slot number $t$. However since we assumed that sensing errors can occur, the value of $X^{(k)}_t$ depends on accuracy of the detector characterized through the measure of two types of errors: on the one hand, detecting a PU on the channel when it is free usually referred to as \textit{false alarm}. On the other hand, assuming the channel free when a PU is occupying it usually referred to as \textit{miss detection}. Let us denote by $\epsilon^{(k)}_n$ and $\delta^{(k)}_n$, respectively the probability of false alarm, and the probability of miss detection characterizing the observation of a channel $n\in\mathcal{D}$ by the $k^{th}$ SU:

$$
\left\{
	\begin{array}{ll} 
	    \epsilon^{(k)}_n=\mathbb{P}\left(X^{(k)}_t=0|S_{a^{(k)}_{t},t}=1\right)\\
	    \delta^{(k)}_n=\mathbb{P}\left(X^{(k)}_t=1|S_{a^{(k)}_{t},t}=0\right)
	\end{array}
\right.
$$     

Finally, the outcome of the sensing process can be seen as the output of a random policy $\pi^{(k)}_{s}(\epsilon^{(k)}_n, \delta^{(k)}_n, S_{a_{t},t})$ such that:
\mbox{$X^{(k)}_{t} = \pi^{(k)}_{s}(\epsilon^{(k)}_n, \delta^{(k)}_n, S_{a_{t},t})$}. The design of such policies \cite{Survey2009} is however out of the scope of this paper.  Depending on the sensing outcome $X^{(k)}_t \in \{0,1\}$, the SU $k$ can choose to access the channel or not.  The access policy chosen in this paper can be described as: ``\textit{access the channel if sensed available}'', i.e. if $X^{(k)}_t=1$. 

%The access strategies depend on the considered scenarios and are discussed later in Section \ref{sec:channel}.

\par Notice that we assume the SUs' detectors to be designed such that for all $k\in\mathcal{K}$ and $n\in\mathcal{D}$, $\delta^{(k)}_n$ (respectively $\epsilon^{(k)}_n$) is smaller or equal to a given interference level allowed by the primary network (respectively, smaller or equal to a given level desired by the SU), although \{$\epsilon^{(k)}_n$, $\delta^{(k)}_n$\} are not necessarily known.  
Moreover, we assume that a packet $D_t=1$ is sent for every transmission attempt. If interference occurs, it is detected and the transmission of the secondary user fails.  Regardless of the channel access policy, when the channel access is granted, the SU receives a numerical acknowledgment. This feedback, usually referred to as \textit{reward} $r^{(k)}_t$ in the Machine Learning literature, informs the SU of the state of the transmission \{succeeded, failed\}. In our scheme, we assume a cooperative scheme is used, i.e. every secondary user shares its reward with the other SUs. All shared information as well as the used communication interface are further discussed in Section \ref{sec:info}.  

%-------------------------------------------------------------------------------------------------
%\subsection{Homogeneous Vs Heterogeneous Environment}
%-------------------------------------------------------------------------------------------------

%----------------------------------Learning Mechanism---------------------------------------------
\section{Learning Mechanism}
\label{sec:lm}
%-------------------------------------------------------------------------------------------------
%-------------------------------------------------------------------------------------------------
\subsection{Joined Resource Allocation-Learning Algorithm}
%-------------------------------------------------------------------------------------------------
\par The learning mechanism aims at exploiting all gathered information to evaluate the most promising resources. Thus, the performance of a learning mechanism highly depends on the sampling model of the rewards (deterministic, stochastic or adversarial for instance). In the case of a stochastic sampling as defined in Section \ref{sec:model_MOSA}, we exploit $UCB_1$ learning mechanisms firstly proposed in~\cite{Poor2008}, since they have proven in~\cite{Jouini2009,Jouini2010}  to be efficient in OSA environments, while having a very low implementation complexity.
\vspace{0.3cm}
\par The estimation of the performance of a resource $n\in\mathcal{D}$ considered by $UCB_1$ indexes relies on the computation of the average reward provided by that resource until the iteration $t$ to which a positive bias is added. The usual form of $UCB_1$ indexes is the following:
\begin{equation}
\label{eq:index1}
B_{T_{n}(t)}=\overline{W}_{T_{n}(t)}+{A_{T_{n}(t)}}
\end{equation}
\noindent where $A_{T_{n}(t)}$ is an upper confidence bias added to the sample mean $\overline{W}_{T_{n}(t)}$ of the resource $n$ after being selected $T_{n}(t)$ times at the step $t$:
\begin{equation}
\label{eq:index2}
\left\{
	\begin{array}{ll} 
		A_{T_{n}(t)}=\sqrt{\frac{\alpha.\ln(t)}{T_{n}(t)}} \\
		\overline{W}_{T_{n}(t)}=\frac{{\sum^{t-1}_{m=0}r_{m}.\textbf{1}_{\{{a}_m=n\}}}}{T_{n}(t)}
	\end{array}
\right.
\end{equation}
For that purpose, we define $B^{(k)}_{T^{(k)}_{n}(t)}$ as the computed index associated to a resource $n$ observed $T^{(k)}_{n}$ times by the $k^{th}$ decision maker until the iteration $t$, and $A^{(k)}_{T^{(k)}_{n}(t)}$ its associated bias.

\vspace{0.3cm}
Let $B(t)$ refer to a $K$ by $N$ matrix such that component of $\{B(t)\}_{\{k,n\}}=B^{(\tilde{k})}_{T_{n}}(t)$, where  $k\in{\mathcal{K}}$, $n\in\mathcal{D}$ and:
$$
\tilde{k}=\left(k-1 + t\right)\oslash K+1
$$
The form of $B(t)$ is explicitly designed to ensure fairness among SUs. As a matter of fact, the rows of $B(t)$ switch at every iteration in a Round Robin way. For the rest of this paper, $B(t)$ is the considered estimated weight matrix for coordination algorithms.

% It is important as some algorithm, such as the Hungarian algorithm, allocates the resources considering the first lines first. Thus if there are several optimal solutions they would always pick the same. The design of $B(t)$ alleviates this problem.
%For the rest of this paper, $B(t)$ is the considered estimated weight matrix for coordination algorithms.
%\vspace{0.3cm}
%\hline
\vspace{0.2cm}
\fbox{\parbox{\textwidth}{
\begin{policy}[$CC-UCB_1(R,\alpha)$] The overall algorithm can be described as follows. 
Let $R$ be a positive integer, $R=1$ if heterogeneous network and $R=K$ if homogeneous network.
\par \textbf{Every $R$ rounds: computation and coordination.}
\begin{itemize}
\item Step 1: Compute $B(t)$ using $UCB_1(\alpha)$ algorithm.
\item Step 2: Compute the output of the coordination mechanism $a^{(k)}_t$ for all users $k$. 

\begin{align}
\max_{\{a^{1}_t,\cdots,a^{K}_t\}}\sum^{K}_{k=1}\{B(t)\}_{\{k,a^{k}_t\}}
\end{align}
%K<n
Thus, every SU is allocated $R$ channels to access in a Round Robin fashion for the next $R$ iteration.
\end{itemize}

\par \textbf{At every iteration during $R$ rounds: sense and access the channels:}

\begin{itemize}
\item Step 3 (for $R$ iterations): Sense the channels and Access them if sensed free.
\end{itemize}

\par \textbf{At the end of $R$ rounds: collaboration-information sharing}
\begin{itemize}
\item Step 4: Share the sensing-access outcomes of the last $R$ rounds.
\end{itemize}
\label{cspolicy:1}
\end{policy}
%\hline
}}
\vspace{0.3cm}

\par As shown by the Channel Selection Policy \ref{cspolicy:1}, the second step relies on a coordination mechanism to perform channel allocation among the SUs. These mechanisms are usually equivalent to \textit{Job Assigment} problems. In the following, we introduce two coordination algorithms in order to allow fair resource allocation among SUs: (\textit{i}) the Hungarian algorithm based coordination  and (\textit{ii}) the Round Robin based coordination.
%\par Although the notion of coordination is one of our approach's corner stones, it has been applied in both Paper \cite{Liu2010} and \cite{Gai2010}. 
%\par In the Appendix \ref{sec:geap} we extensively introduce such coordination problems and introduce the considered coordination mechanism \ref{coord_2} (Round Robbins based coordination) and coordination mechanism \ref{coord_1} (Hungarian algorithm based coordination) of this paper. These mechanism are slightly modified to allow fair resource allocation among users. These modifications are considered as part of the contributions of this work. 

%-------------------------Decision Making Framework----------------------------------------------
\section{General Resource Allocation Problem}
%------------------------------------------------------------------------------------------------
\label{sec:geap}
%One of the contributions of this paper, is to introduce a general resource allocation framework to discuss OSA scenarios. In this Section, we emphasize that OSA related problems are simple applications of this general framework. Moreover the coordination mechanism considered for the theoretical an illustrative Sections are introduce within this section.
%However, the reader do not need to fully understand these mechanisms to follow our theoretical analysis as well as the empirical evaluation. Consequently, in order to avoid overwhelming the reader, we chose to discuss coordination in the Appendix rather than in a common section.   
%-------------------------------------------------------------------------------------------------
\subsection{Coordination and Job Assignment Problems}
\label{subsec:cord}
%-------------------------------------------------------------------------------------------------
\par We argue in this subsection that the coordination of multi-secondary users can be formulated as a job assignment problem. We first introduce the general notations related to the job assignment framework. Then we present this latter as an adequate tool to model OSA related coordination problems.
\vspace{0.3cm}
\par Let us consider a set $\mathcal{K}$ of $K$ \textit{workers} or \textit{decision makers} and a set $\mathcal{D}$ of $N$ \textit{jobs} or \textit{resources}. Let us denote by $\lambda$ the $K$ by $N$ weight (or cost) matrix where $\{\lambda\}_{\{k,n\}}=\lambda^{(k)}_n$  refers to a weight associate to the \textit{decision maker} $k\in\mathcal{K}$ assigned to the \textit{job} or \textit{resource} $n\in\mathcal{D}$ such that:
$$
\lambda=
\begin{bmatrix}
\lambda^{(1)}_1&\cdots&\lambda^{(1)}_N\\
\lambda^{(2)}_1&\cdots&\lambda^{(2)}_N\\
 &\cdots& \\
\lambda^{(K)}_1&\cdots&\lambda^{(K)}_N
\end{bmatrix}
$$
\par We assume that every \textit{decision maker} can be assigned on a unique \textit{resource}. Moreover, every \textit{resource} can be handled by only one \textit{decision maker}. Let $a_n\in\mathcal{K}$ refer to the assigned \textit{decision maker} to the \textit{resource} $n\in\mathcal{D}$. The resource allocation problem can be formalized as follows. Find an optimal set of assignments such that the total weight is maximized (or equivalently, the total cost minimized):
\begin{align}
\label{eq:coord}
\max_{\{a_1,\cdots,a_N\}}\sum^{N}_{n=1}\lambda^{(a_n)}_{n}\textbf{1}_{\{\exists \ a_n\}}
\end{align}
\noindent where the logic expression\footnote{Indicator function: 
  $\textbf{1}_{\{logical\_expression\}}$=\{{1 if logical\_expression=true} ; {0 if logical\_expression=false}\}.} $\{\exists \ a_n\}$ refers to the existence of a \textit{decision maker} assigned to the resource $n$.

\vspace{0.3cm}  
\par In the case of OSA a coordinator generally aims at canceling harmful interference among the  SUs. To that purpose, a coordinator usually allocates different resources to different users, or uses advanced signal processing techniques to alleviate interference effects on the users' performances (e.g., Time Division Multiple Access, Frequency Division Multiple Access or Code Division Multiple Access to name a few): 

\begin{definition}[Coordinator or Facilitator]
Let $\mathcal{K}$ refer to a set of decision making agents. We refer to as \textit{Coordinator} or \textit{Facilitator} any real or virtual entity that enables the different \textit{decision makers} to jointly plan their decisions at every iteration. \\
\end{definition}

\par For the sake of coherence in speech, let us consider a set $\mathcal{K}$ of $K$ SUs (viz., the workers or decision makers) willing to exploit a set $\mathcal{D}$ of $N$ primary channels (viz., the resources). Moreover let $\{\mu_n\}_{\{n\in\mathcal{D}\}}$ and $\pi^{(k)}_{s}$ denote, respectively, a characteristic measure that quantifies the quality of the primary channels (e.g., their expected availability or Signal to Noise Ratio for instance) and a sensing policy that characterizes the observation abilities of the $k^{th}$ SU. 
Then $\lambda^{(k)}_n=f_{\pi^{(k)}_{s}}\left(\mu_n\right)$  represents the quality of a primary resource observed by the $k^{th}$ SU, where $f\left(\cdot\right)$ represents a (possibly implicit) functional relationship that relates primary resources' quality to SUs observations. Consequently, the stated problem in Equation \ref{eq:coord} is equivalent, when allocating primary resources among SUs, to maximize the secondary network's observed performance. 

%-------------------------------------------------------------------------------------------------
\subsection{Coordination Mechanisms based on The Hungarian Algorithm}
\label{subsec:ha}
%-------------------------------------------------------------------------------------------------
\par Suggested in 1955 by H. W. Kuhn \cite{Kuhn1955}, the Hungarian method is a matching algorithm that solves the job assignment problem in polynomial time. It mainly takes as an input the matrix $\lambda$ (or its opposite, depending on whether it is a maximization or minimization approach) and provides as an output a binary matrix that contains a unique $1$ per row and per column. This output indicates the resource allocation to the workers. 
\par Many assignment combinations can verify the stated problem in Equation \ref{eq:coord}. The Hungarian algorithm provides one solution among the set of optimal matching solutions. This solution mainly depends on the matrix $\lambda$. Inverting two columns can lead to a different optimal solution if such solution exists. It is thus necessary to consider, for fairness reasons among SUs, a permutation mechanism that changes the order of the rows of the weight matrix at every new iteration $t$. To this goal,  we introduce  the following coordination algorithm:
\vspace{0.2cm}
\begin{coordination}[Hungarian Algorithm based Coordination]
\label{coord_2}
Let $t=0,1,2,\cdots$ refers to a discrete sampling time and let $\{\lambda^{(k)}_n(t)\}_{n\in\mathcal{D}}$ refers to weights associated to the decision maker $k\in\mathcal{K}$ at the iteration $t$.
Let $\lambda(t)$ refer to a $K$ by $N$ matrix such that  $\{\lambda(t)\}_{\{k,n\}}=\lambda^{(\tilde{k})}_n(t)$,  $k\in{\mathcal{K}}$, $n\in\mathcal{D}$ and:
$$
\tilde{k}=\left(k-1 + t\right)\oslash K+1
$$
\noindent where $a \oslash b$ refers to the modulo operator that returns the remainder of the division of $a$ by $b$.

Let $H(t)$ refer to the output of the Hungarian algorithm with input $\lambda(t)$. 
\par Then the $k^{th}$ decision maker is assigned the resource $a^{(k)}_t$ verifying:
\begin{align}
a^{(k)}_t=n \ s.t. \ H(t)_{\{\tilde{k},n\}}=1;
\end{align}
\end{coordination}
\vspace{0.3cm}

%-------------------------------------------------------------------------------------------------
\subsection{Coordination Mechanisms based on Round Robin Algorithm}
\label{subsec:rr}
%-------------------------------------------------------------------------------------------------
We consider in this subsection, a particular case of the introduced job assignment problem: Symmetric workers.
\begin{definition}[Symmetric Behavior]
Let $\mathcal{K}$ refer to a set of decision making agents. These agents are said to have a \textit{Symmetric Behavior} if their optimization criteria, their communication abilities as well as their decision making policies are the same. In OSA contexts, a network with Symmetric Behavior transceivers is thus referred to as a \textit{Symmetric Network}.\\
\end{definition}

\par This can be formalized as particular weight matrix with the same rows for all $k \in \mathcal{K}$, i.e., let $n$ be a resource, $n\in\mathcal{D}$ then:
$$
\forall k\in\mathcal{K} \ \{\lambda\}_{\{k,n\}}=\lambda_n
$$

In this context a very simple coordination algorithm can ensure fairness among workers\footnote{Although the suggested form is original, the coordination algorithm is a simple Round Robin allocation scheme. It has been already suggested in \cite{Liu2010} in an OSA context without considering collaboration.}:
\begin{coordination}[Circular Coordination (Round Robin)]
\label{coord_1}
Let $t=0,1,2,\cdots$ refers to a discrete sampling time. We define $t'=\{0, 1, \cdots, \left\lfloor t/K\right\rfloor \}$ as a sequel of integers updated every $K$ iterations.
Let $\{\lambda(t)\}$ refer to the weight matrix computed at the iteration $t$, and
let $\sigma_n(t)$ be the permutation function used at the iteration $t$ to order the rows of the weight matrix values.
We assume that $\{\lambda(t)\}$ and $\sigma_n(t)$ are computed every $K$ iterations such that for all $t \in [Kt', K(t'+1)-1]$, $\{\lambda(t)\}=\lambda(Kt')$ and $\sigma_n(t)=\sigma_n(Kt')$.
\par Then the $k^{th}$ decision maker selects the channel $n$ verifying:
\begin{align}
\left\{
	\begin{array}{ll}
\sigma_n(t)=a^{(k)}_t\\
a^{(k)}_t=\left(k-1 + t \right)\oslash K+1
\end{array}
\right.
\end{align}
\end{coordination}
\vspace{0.3cm}
\par Coordination algorithm \ref{coord_1} needs to know that the network is perfectly symmetric. In case this knowledge is unavailable or the network is non-symmetric, this coordination scheme could fail. Moreover, in real scenarios, the weight matrix is usually unknown and every worker can solely access one row of the matrix, i.e. the one related to his own perception of the environment (usually prone to detection errors). Consequently, OSA related problems appears as Job Assignment problems under uncertainty. 
\par Thus, we suggest in this paper to introduce collaboration and coordination based learning mechanisms among  SUs in order to alleviate the lack of information at each SU  and to converge to the optimal resource allocation. In order to compute an estimation of the weight matrix, we assume a collaboration behavior among workers to share information (discussed in Section \ref{sec:info}). The shared information enables a learning mechanism to compute the estimated quality matrix as described in Section \ref{sec:lm}. The performance of Coordination algorithm \ref{coord_2} is empirically analyzed in Section \ref{sec:simu}, while, in the case of symmetric networks, the performance of Coordination algorithm \ref{coord_1} is theoretically analyzed in Section \ref{sec:theory}.

%---------------------------------------Theor-----------------------------------------------
\section{Theoretical analysis }
\label{sec:theory}
%-------------------------------------------------------------------------------------------
\subsection{Definitions of the Reward and the Expected Cumulated Regret}
%-------------------------------------------------------------------------------------------------
Since we consider a coordinated network, it reasonable to assume that interference among SUs is null. Thus, relying on the previously introduced notations and assumptions, the throughput achieved by a $SU_k$, $k \in \mathcal{K}$, at the slot number $t$ can be defined as:

\begin{equation}
\label{eq:rewards}r^{(k)}_t=S_{a^{(k)}_{t},t}X^{(k)}_{t}
\end{equation}
\noindent which is the reward considered in this particular framework, where $r^{(k)}_t$ equals $1$ only if the channel is free and the SU observes it as free. Consequently, the expected reward achievable by a given secondary user $SU_k$ using a channel $a^{(k)}_{t} \in \mathcal{D}$ can be easily computed:
\begin{align}
\label{Eq:gain}
\mathbb{E}\left[r^{(k)}_t\right]=\mathbb{P}\left(X^{(k)}_{t}=1|S_{a^{(k)}_{t},t}=1\right)\mathbb{P}\left(S_{a^{(k)}_{t},t}=1\right)
\end{align}
\noindent which equals to, in this case:

\begin{align} 
\mathbb{E}\left[r^{(k)}_t\right]=\left(1-\epsilon^{(k)}_{a^{(k)}_{t}}\right)\mu_{a^{(k)}_{t}}
\end{align}
\par To relate with the job assignment problem described in Section \ref{sec:geap}, the weight matrix $\lambda$ is equal, is this context, to the matrix $\{\mathbb{E}\left[r^{(k)}_t\right]\}_{\{k\in\mathcal{K},n\in\mathcal{D}\}}$.
\par We usually evaluate the performance of a user $k$ by its expected cumulated throughput after $t$ slots defined as:

%\begin{align}
\begin{equation}\mathbb{E}\left[W^{(k)}_t\right]=\mathbb{E}\left[\sum^{t-1}_{m=0}r^{(k)}_m\right]\end{equation}
%\end{align}
%\noindent where the suffix $\pi$ is used to emphasize that the CA uses the channel selection policy $\pi$. 
 Notice that in this case, $r^{(k)}_t$ follows a Bernoulli distribution.
 % Several other algorithms are possible to answer this distribution within MAB problems such as Robbins or Agrawal's index policies \cite{Robbins1985,Agrawal1995}. They would however fail in more complex scenarios where $r^{(k)}_t$ follows an unknown distribution in $[0,\ 1]$ for instance. As a matter of fact, they need the knowledge of the exact reward distribution, whereas $UCB_1$ guaranties order optimality for any bounded reward distribution (even unknown). This briefly justifies the choice of $UCB_1$.
An alternative representation of the expected performance of the learning mechanism until the slot number $t$ is described through the notion of \textit{regret} $R^{(k)}_t$ (or expected regret of the $SU_k$). The regret is defined as the gap between the maximum achievable performance in expectation and the expected cumulated throughput achieved by the implemented policy.  

\begin{equation}
\label{eq:regret}	R^{(k)}_t=\sum^{t-1}_{m=0}\max_{a^{(k)}_{t}\in\mathcal{A}^{(k)}_t}\mathbb{E}[r^{(k)}_t]-\mathbb{E}\left[W^{(k)}_t\right]
\end{equation}
\noindent where $\mathcal{A}^{(k)}_t$ denotes the subspace of channels that a given $SU_k$ can access at the slot time $t$, $\mathcal{A}^{(k)}_t\subseteq\mathcal{D}$.

%-------------------------------------------------------------------------------------------------
\subsection{Theoretical Results: Symmetric Networks}
%-------------------------------------------------------------------------------------------------
In Symmetric Networks, the expected quality of a channel $n$ observed by all SUs is the same:  $\forall k\in\mathcal{K} \ \lambda^{(k)}_n=\lambda_n $. If the symmetry property is known to SU, all collected information on the probed channels at the slot number $t$ is relevant to every SU. Thus, it can be used to improve their overall learning rate. As matter of fact, in this context, the SUs combine at every iteration all gathered rewards into one common information vector $i_t$ such that $i_t=\{i_{t-1}, \{a^{(k)}_t, r^{(k)}_t\}_{k\in\mathcal{K}}\}$. Hence, the UCB indexes computed by the SUs at every slot number $t$ are also the same, i.e., for all users $k\in\mathcal{K}$, $B^{(k)}_{T^{(k)}_{n}}(t)=B_{T_{n}}(t)$. Notice that in Symmetric Networks the optimal set of channels $\mathcal{D}^*$ is composed of the $K$ channels with the highest expected reward. Consequently a simple Round Robin based coordination algorithm, as described in Coordination \ref{coord_2} is optimal (avoids harmful interference and is fair).

\par In the next Theorem, we show that the regret of the $k^{th}$ SU in a Coordinated and Collaborative Symmetric Network is upper bounded by a logarithmic function of the number of iterations $t$.
\vspace{0.2cm}
\par \fbox{\parbox{\textwidth}{
\begin{theorem}[Upper Bound of the Regret]
\label{thm1}
{Let us consider $K\geq1$ Symmetric Secondary Users and $N\geq K$ Primary channels. The SUs are assumed to have limited observation abilities defined by their  parameters $\{\epsilon_n,\delta_n\}$ for every channel $n$. Assuming that the Secondary Network follows the Coordination Policy \ref{coord_1} to select and access the primary channels, relying on $UCB_1$ algorithm with parameter $\alpha>1$, then every SU suffers an expected cumulated regret $R^{(k)}_t$, after $t$ slots, upper bounded by a logarithmic function of the iteration $t$:}
\begin{align}
R^{(k)}_t \leq \sum_{n\notin \mathcal{D}^{*}}\frac{4\alpha\left(\bar{\lambda}^{*}-\lambda_{n}\right)}{K\Delta^2_n}\ln\left(t+K-1\right)+o\left(\ln(t)\right)
\label{regret}
\end{align}
\noindent where the following notations were introduced: 
$$
\left\{
	\begin{array}{ll}  
			\lambda_n=\left(1-\epsilon_n\right)\mu_{n}\\
			\bar{\lambda}^{*}=\frac{\sum_{n\in\mathcal{D}^{*}}\lambda_n}{K}\\ 
  		\Delta_n=\min_{n\in\mathcal{D}^{*}}\{\lambda_n\}-\lambda_{n}
	\end{array}
\right.
$$
\end{theorem} 
}}
\vspace{0.3cm}
\begin{proof}
This proof relies on two main results stated and proven in Lemma \ref{lemme_regret1} and Lemma \ref{lemme_sample} (C.f. Appendix).
As a matter of fact, Lemma \ref{lemme_regret1} shows that the regret can be upper bounded by a function of the expected number of pulls of sub-optimal channels:
\begin{align}
R^{(k)}_t \leq \sum_{n\notin \mathcal{D}^{*}}\frac{\left(\bar{\lambda}^{*}-\lambda_{n}\right)\mathbb{E}\left[T_n\left(\left\lfloor \frac{t}{K}\right\rfloor K+K-1\right)\right]}{K}
\end{align}
Then Lemma \ref{lemme_sample} upper bounds $\mathbb{E}\left[T_n\left(\left\lfloor \frac{t}{K}\right\rfloor K+K-1\right)\right]$ by a logarithmic function of number of iterations $t$:
\begin{align}
\mathbb{E}\left[T_n\left(\left\lfloor \frac{t}{K}\right\rfloor K+K-1\right)\right] \leq \frac{4\alpha}{\Delta^2_n}\ln\left(t+K-1\right)+o\left(\ln(t)\right)
\end{align}
\end{proof}
\par For the case $K=1$, $\epsilon_n=\epsilon$ and $\delta_n=\delta$, we find the classic result stated in \cite{Jouini2011}:
\begin{align}
R_t \leq \sum_{n\neq1} \frac{4\alpha}{\left((1-\epsilon)(\max_{n\in\mathcal{D}}\{\mu_n\}-\mu_n)^2\right)}\ln\left(t\right)+o\left(\ln(t)\right)
\end{align}

%-------------------------------------------------------------------------------------------------
\subsection{Theoretical Results: Non-Symmetric Networks}
%------------------------------------------------------------------------
\par In the case of Non-Symmetric Networks, we can apply the upper bound provided in~\cite{Gai2010bis}. As a matter of fact, our approach that decomposes, on the one hand the learning step and on the other hand the coordinating step, is equivalent to the algorithm referred as \textit{Learning with Linear Regret} (LLR) in~\cite{Gai2010bis}. More specifically, the authors of~\cite{Gai2010bis} prove that, if the exploration parameter of the $UCB_1$ algorithm, i.e. the $\alpha$ factor, verifies this condition: $\alpha\geq L$ where $L=N\wedge K=K$ where $\wedge$ refers to the minimum operator, then the LLR algorithm has an order optimal behavior (i.e., expected cumulated regret upper bounded by a logarithmic function of the time). In our case, the logarithmic regret scales linearly with the value: ${(N\wedge K)^3NK}$ as reported in \cite{Gai2010bis}. 
\par However fairness is not consider in~\cite{Gai2010bis}. Our suggested joint coordination-learning mechanism alleviates this problem. It is easy to verify that the same results discussed in~\cite{Gai2010bis} hold also when when the Coordination algorithm \ref{coord_2} is used for spectrum selection. Consequently, a joined coordination-learning mechanism in Non-Symmetric environments is order optimal. 
\par Although this result is fundamental to many resource allocation problems under uncertainty,  two questions remain unanswered in~\cite{Gai2010bis}: 
\begin{itemize}
\item With the result provided for Non-Symmetric Environment, it is obvious that the same mechanisms would also work for Symmetric Environments. Is it possible to provide tighter bounds for the regret and to use smaller value for the exploration parameter $\alpha$?
\item Although the theory constrains $\alpha$ to values larger than $K$ (in our case), does it mean that the algorithm fails for smaller values? Notice that the larger $K$ is, the longer it takes to converge. 
\end{itemize}
Both questions are tackled in this paper. On the one hand, the previous subsection tackled the first question. We see from the results of Theorem \ref{thm1} that the logarithmic function scales as $1/K$, improving tremendously the scale found in the case of heterogeneous environments. On the other hand, the simulations discussed in Section \ref{sec:simu} suggest a piece of answer to the second question.

%------------------------Information and cooperation among Secondary Users}---------------------------
\section{Information Sharing: Discussion}
%-----------------------------------------------------------------------------------------------------
\label{sec:info} 
%We detail in this section the exchanged information and  its purpose.
%-----------------------------------------------------
%\subsection{Common Control Channel}
%-----------------------------------------------------
\par An efficient communication process relies on reliable information exchange. Thus, we assume in this paper that the communication interface used by Cognitive Radio (CR) SUs to share information is a Common Control Channel\footnote{Whether to use or not CCCs for cognitive radio networks is still a matter of debate in the CR community. This debate is however out of the scope of this paper. Notice that the conclusions of this study would still apply if we assumed any other kind of reliable information exchange interface among secondary users.} (CCC). The CCC is used, on the one hand, between a transmitter and a receiver (which can be a secondary base station or another SU), and on the other hand, among all transmitters and receivers for cooperation purposes. The information transmitted through this vessel is furthermore assumed to be received without errors. 
\par Thus from a \textit{Transmitter-Receiver}'s perspective, the purposes of the CCC are twofold: configuration adaptation and acknowledgment messages transmission.
\subsubsection{Configuration adaptation} To initiate a transmission, both the transmitter and the receiver have to agree on a particular frequency band and on a communication configuration (e.g., modulation). In this particular case, \textit{configuration} refers, solely, to \textit{frequency band}. Thus we assume that at every slot $t$ the transmitter informs the receiver of the channel selection outcome before transmitting.
\subsubsection{Acknowledgment} At the end of every transmission attempt the receiver has to confirm the reception of the transmitted parquet. In case of a successful transmission, the transmitter receives an \textit{ACK} message from the receiver. Otherwise, in case of PU interference, it receives a \textit{NACK} message. 
\subsubsection{Information sharing} As mentioned in Section \ref{subsec:SU}, at the end of every slot $t$, and for cooperation purposes, a communication period is dedicated to share computed rewards information among SUs.  As a consequence, a given SU can coordinate its behavior according to other SUs. Moreover, each SU can learn faster the spectrum occupation by relying on the outcomes of the other SUs' attempts, gathered on bands it did not address.

%----------------------------------Simulation-----------------------------------------------
\section{Empirical Evaluation: Simulation Results}
\label{sec:simu}
%------------------------------------------------------------------------------------------- 
% \textcolor{red}{To fill once all subsections are written.}
In this section, we describe and show the simulation results  aimed at illustrating the herein suggested resource selection mechanisms. 
We first describe the general experimental protocol and the considered scenarios in Subsection \ref{subsec:protocol}. Subsection \ref{subsec:simu} presents and discusses the simulation results pertaining to the regret analysis. Subsection \ref{subsec:simuMarco} show the results pertaining to the secondary network performance analysis.
%------------------------------------------------------------------------------------------- 
\subsection{Scenario and experimental protocol for the regret analysis}
\label{subsec:protocol}
%------------------------------------------------------------------------------------------- 

\par We consider $3$ secondary users willing to exploit $10$ primary channels with unknown expected occupancy patterns  $\mu=\{\mu_n\}_{\{1,\cdots,10\}}$.
For the sake of generality, we do not provide explicit numerical values to PUs' channel occupancy and to the probability of false alarms. The impact of sensing errors has been analyzed and illustrated in a previous work \cite{Jouini2011}.
\par We denote by $\lambda^{(k)}_n$ the expected reward of a resource $n$ observed by a user $k$. We, however consider that the occupation state $n$ observed by a user $k$ at the slot $t$ follows a Bernoulli distribution with parameter $\lambda^{(k)}_n$. Thus, the application to OSA related scenarios is straightforward as: $\lambda^{(k)}_n=\left(1-\epsilon^{(k)}_n\right)\mu_n$ in this context.

\par For illustration purposes we tackle two scenarios. On the one hand we consider $3$ symmetric users. While on the other hand, we consider that the $3$ secondary users are divided into $2$ sets: two symmetric users sharing the spectrum with a last secondary user whose optimal channel do not belong to the set of optimal channels of the other secondary users, such that:
\begin{scenario}[Symmetric network]
\label{scenario1}
We consider a quality matrix $\lambda$ defined as:
$$
\lambda=\begin{bmatrix}
0.1&0.1&0.2&0.3&0.4&0.5&0.6&0.7&0.8&0.9\\
0.1&0.1&0.2&0.3&0.4&0.5&0.6&0.7&0.8&0.9\\
0.1&0.1&0.2&0.3&0.4&0.5&0.6&0.7&0.8&0.9
\end{bmatrix}
$$
\end{scenario}
\begin{scenario}[Non-symmetric network]
\label{scenario2}
We consider a quality matrix $\lambda$ defined as:
$$
\lambda=\begin{bmatrix}
0.1&0.1&0.2&0.3&0.4&0.5&0.6&0.7&0.8&0.9\\
0.1&0.1&0.2&0.3&0.4&0.5&0.6&0.7&0.8&0.9\\
0.1&0.1&0.2&0.3&0.4&0.7&0.9&0.7&0.7&0.6
\end{bmatrix}
$$
\end{scenario}
These scenarios aim at illustrating both Hungarian and Round Robin based coordination algorithms. We expect the channel selection algorithm, relying on both learning and coordinations mechanisms to be able to converge to the set of optimal channels in Scenario \ref{scenario1}. However, in Scenario \ref{scenario2} only the Hungarian algorithm based coordinator is illustrated as a Round Robin approach would be inefficient. 
\par During all experiments, the learning parameter $\alpha$ is selected such that $\alpha=1.1$ (to respect the conditions of Theorem \ref{thm1}). Notice that these simulations were conducted so as their respective results and conclusion could be generalized to more complex scenarios. 
\par Finally, the presented results are averaged over $30$ experiments with a final horizon equal to $1\ 000\ 000$ slots to obtain reliable results.

%------------------------------------------------------------------------------------------- 
\subsection{Simulation results: Regret Analysis}
\label{subsec:simu}
%------------------------------------------------------------------------------------------- 
 
\par
\begin{figure*}[t!]
\centerline{
\subfigure[\label{regret1}] {\includegraphics[width=0.5\textwidth, height=7cm]{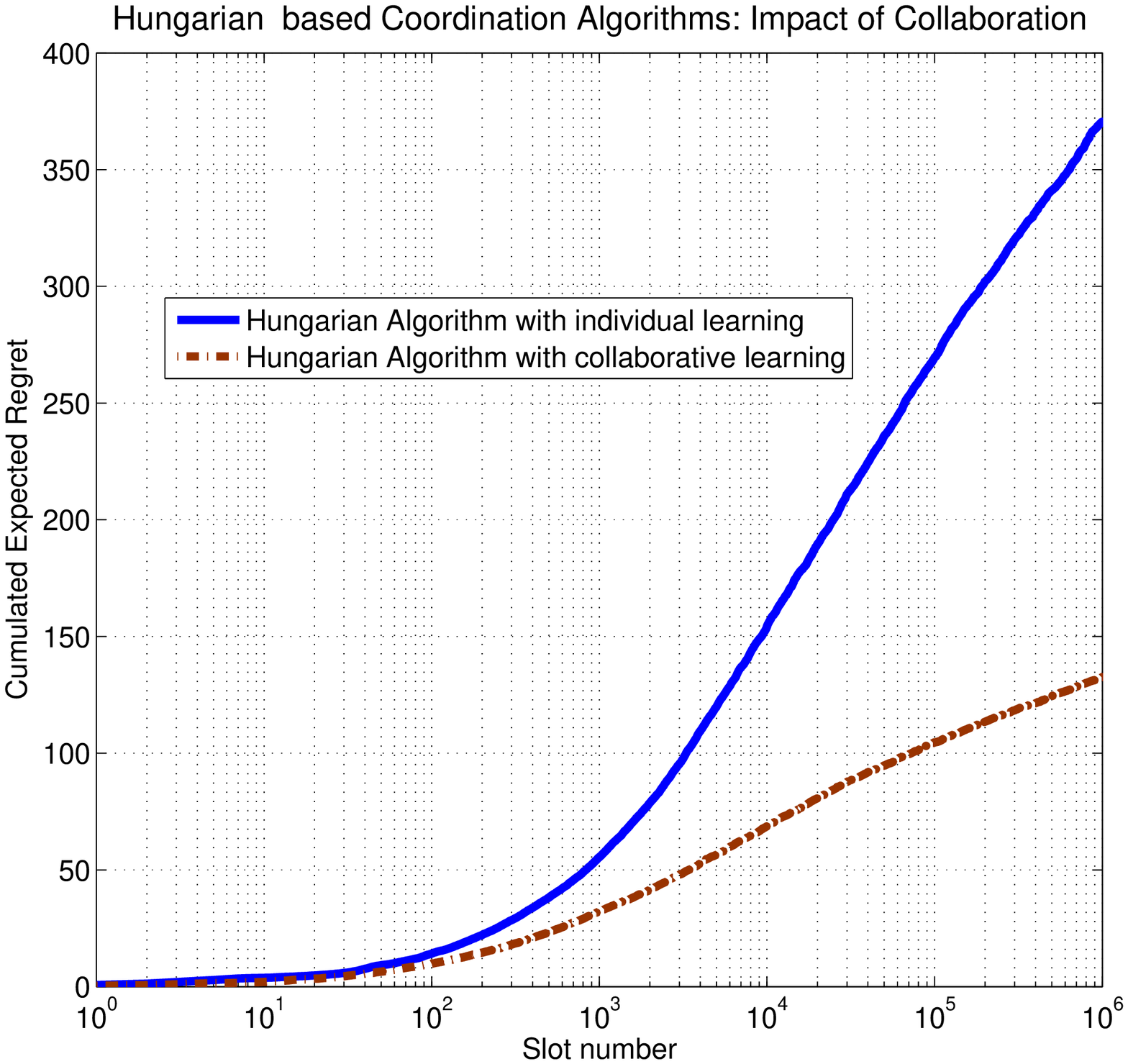}}
\subfigure[\label{regret2}] {\includegraphics[width=0.5\textwidth, height=7cm]{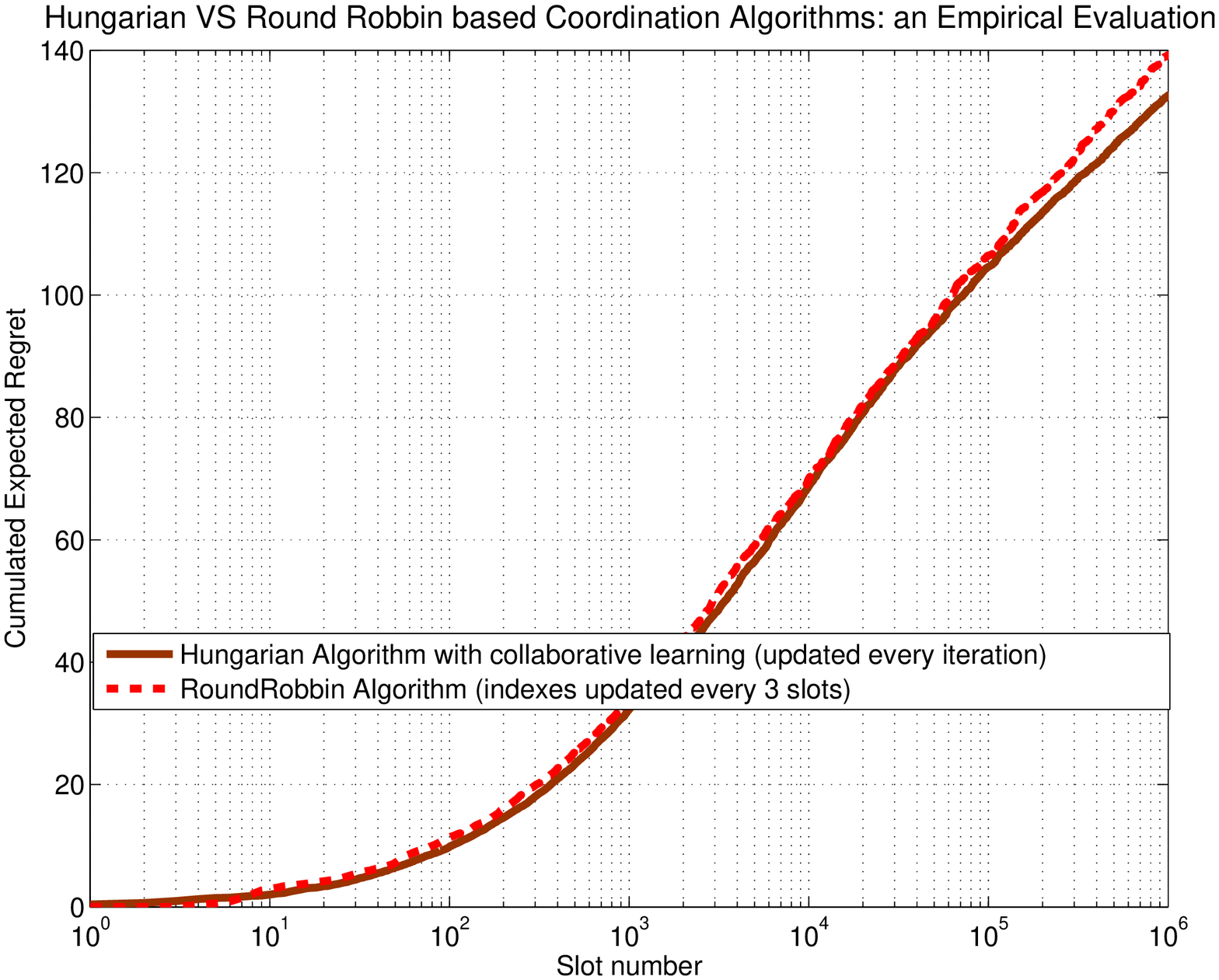}}
}
 \caption{Collaboration, Learning and Coordination in the case of Symmetric Networks: averaged regret. The simulation results show that both Hungarian algorithm and Round Robin based coordinators can efficiently learn to allocate the resources among the SUs. All curves are computed with $\alpha=1.1$. Left Figure shows the impact of collaboration on the learning process in symmetric networks. Right curves compares learning mechanisms with both Hungarian coordination or Round Robbin coordination. We notice that their performance is quite similar.   }
\end{figure*}

 The averaged regret -over the number of SUs- of four algorithms are illustrated in Figures \ref{regret1}  and \ref{regret2} in the context of Scenario \ref{scenario1}: Figure \ref{regret1} shows the regrets of the Hungarian algorithm, respectively, with or without common information vector (i.e. with or without collaborative learning), while Figure \ref{regret2} correspond to Round Robin based coordination algorithms with common information vector. In this latter case, one algorithm updates its information vector every $3$ iterations (i.e., every $K$ iterations as considered in Theorem \ref{thm1}), while the second one updates its information vector every slot.
\par  On the one hand, Figure \ref{regret2} illustrates Theorem \ref{thm1}. As a matter of fact, we observe that the regrets of Round Robin based algorithms are  similar and have indeed a logarithmic like behavior as a function of the slot number. This behavior is observed for all four simulated algorithms. Secondly, as expected, the Hungarian based coordinator with collaborative learning performs as well as Round Robin based coordinators.
\par On the other hand, Figure \ref{regret1} shows the impact of coordination with individual learning (the shared information is only used for coordination purpose). In this case the regret grows, as expected, larger by a factor approximatively equal to $K$. In this case where the users are symmetric but unaware of that fact, they do not exploit other users' information to increase their respective learning rate. The collected information from their neighbors is solely used to compute the quality matrix $\lambda$ to enable coordination. Thus, we observe in Figure \ref{regret1} that the Hungarian algorithm is still able to handle it however, as already noticed, with a loss of performance. 

\par
\begin{figure}
	\centering
		\includegraphics[width=0.8\textwidth, height=7cm]{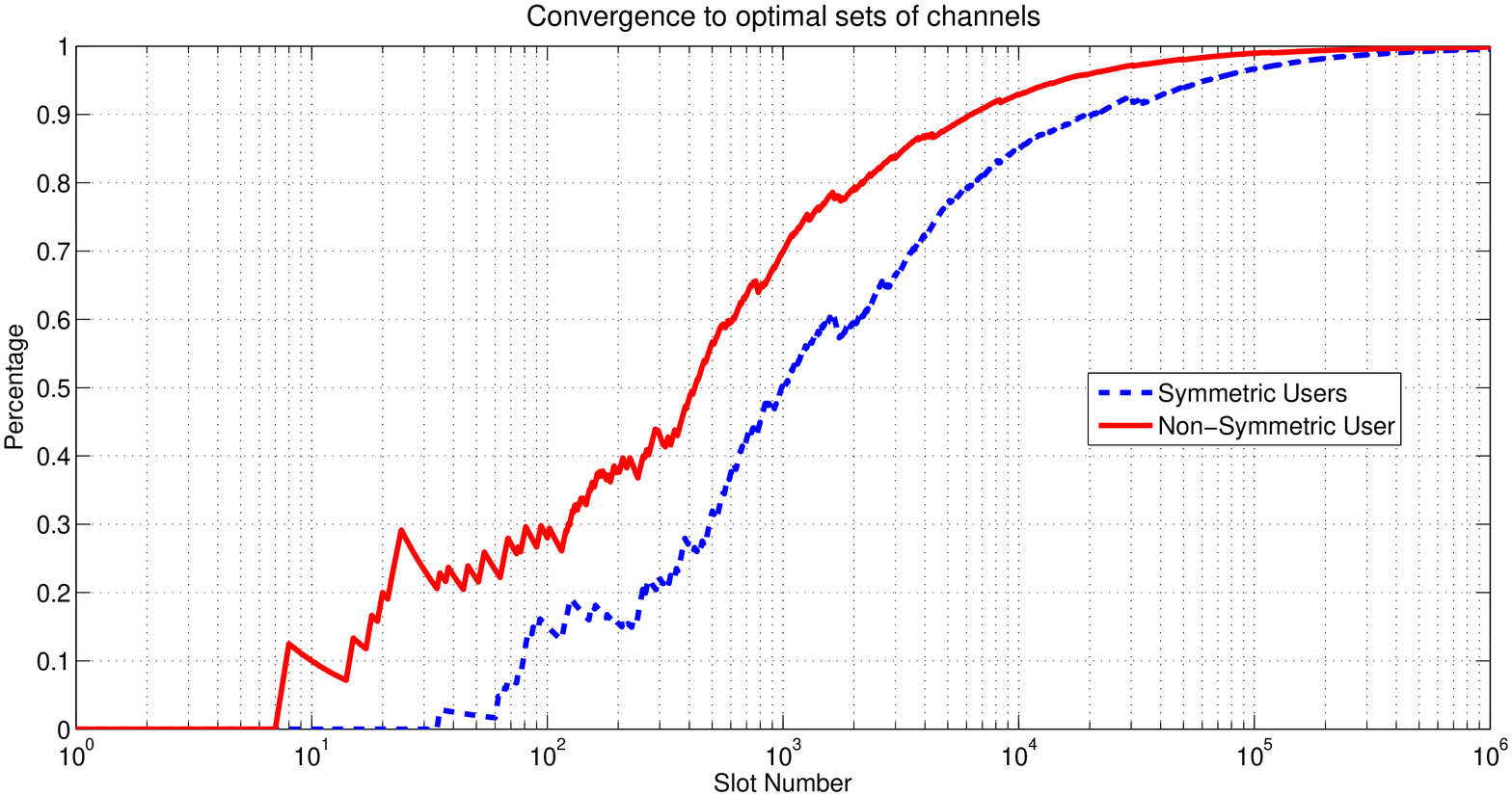}
	\caption{Percentage of time the Hungarian algorithm based coordinator allocates the different secondary users to their respective optimal sets. The exploration parameter $\alpha$ is chosen equal to $1.1$. This value is smaller than the minimum value suggested by the theory. We observe however that the algorithm remains consistent.}
		\label{fig_convergence}
\end{figure}

 In the case of Scenario \ref{scenario2}, Round Robin based coordination algorithms are in general not efficient. Consequently, we do not illustrate them in this context. Figure \ref{fig_convergence} shows the proportion of time the Hungarian algorithm based coordinator allocates the different secondary users to their respective optimal sets. We can observe that the curves increase rather quickly which indicates that the algorithm allocates the SUs to their respective optimal sets most of the time after a first learning phase. Theoretical analysis as well as testbed-based experiments are currently under investigation to confirm these results.

%\begin{figure}
%\includegraphics[width=0.5\textwidth, height=7cm]{regretMulti_usersHun.eps} \hfill
%\includegraphics[width=0.5\textwidth, height=7cm]{regretMulti_usersHvsRR.eps}\\
%\caption{Collaboration, Learning and Coordination in the case of Symmetric Networks: averaged regret. The simulation results show that both Hungarian algorithm and Round Robin based coordinators can efficiently learn to allocate the resources among the SUs. All curves are computed with $\alpha=1.1$. Left Figure shows the impact of collaboration on the learning process in symmetric networks. Right curves compares learning mechanisms with both Hungarian coordination or Round Robbin coordination. We notice that their performance is quite similar. }
%\label{fig_regret}
%\end{figure}

%---------------------------------------Added by Marco-----------------------------------------------

\subsection{Simulation results: Network Performance Analysis}
\label{subsec:simuMarco}
In this subsection, we evaluate the performance of joint collaboration-cooperative learning scheme from the point of view of secondary network performance. To this aim, we model a primary network with $N$=10 channels, and a secondary network composed of $K$=4 transmitter nodes. The temporal occupation pattern of the $N$ channels is defined by this vector $\theta$ of Bernoulli distributions: $\{$0.1, 0.1, 0,2, 0.3, 0.4, 0.5, 0.6, 0.7, 0.8, 0.9$\}$. All the $SU$ have a fixed probability of sensing miss-detection and sensing  false-alarm, i.e. a Symmetric Network scenario is considered. Unless specified otherwise, we set  $\epsilon^{(k)}_n$=0.2, for all SU $k$ and channel $n$. At each slot, each SU $k$ decides a channel to sense, and transmits a packet of $1000$ bytes if the channel is found idle. No transmission attempt is performed in case the channel is sensed occupied by a PU. Both interferences among SUs  and between a SU and a PU are taken into account in the model. If no interference occurs during the SU transmission, then an ACK message is sent back  to the SU transmitter. Otherwise, the data packet is discarded by the SU receiver node.  Thus, at each slot $t$, each SU $k$ can experience a local throughput $TP^k(t)$ equal to 0 or 1000 bytes, based on interference and sensing conditions. The average network throughput $NTP(t)$ is defined as the average amount of byte successfully transmitted in the secondary network at each slot $t$, i.e.: $NTP(t)$=$E[\sum_{k=1}^{K} TP^k(t)]$.\\
We consider four different configurations of learning, cooperation and coordination schemes in our analysis:
\begin{itemize}
\item C1 (\textit{Random, No Learning}):  no learning is employed by SUs. At each slot, each SU chooses randomly the channel to sense among the available $N$ channels.  
\item C2 (\textit{Individual Learning, No Coordination}):  each SU employs the $UCB_1$ algorithm to learn the temporal channel usage. No coordination and collaboration mechanisms are used. At each slot $t$, each SU $k$ chooses randomly based on the local $UCB_1$-index associated to each channel. More specifically, the probability to select channel $n$ is computed proportional to $1- B^{(k)}_{T^{(k)}_{n}(t)}$. The probabilities are normalized so that there value is between $0$ and $1$, and their sum equals one.
\item C3 (\textit{Cooperative Learning, No Coordination}): as before,  each SU employs the $UCB_1$ algorithm to learn the temporal channel usage, and shares the rewards received at each slot $t$. However, no collaboration mechanism is used. The channel selection is performed as the previous case. 
\item C4 (\textit{Cooperative Learning, Cooperation}): the complete Channel Selection Policy 1 described in Section \ref{sec:lm} is evaluated. The Round Robin algorithm is considered for channel access coordination. 
\end{itemize}
\par
\begin{figure*}[t!]
\centerline{
\subfigure[\label{thr1}] {\includegraphics[height = 7cm,
width=8.5cm]{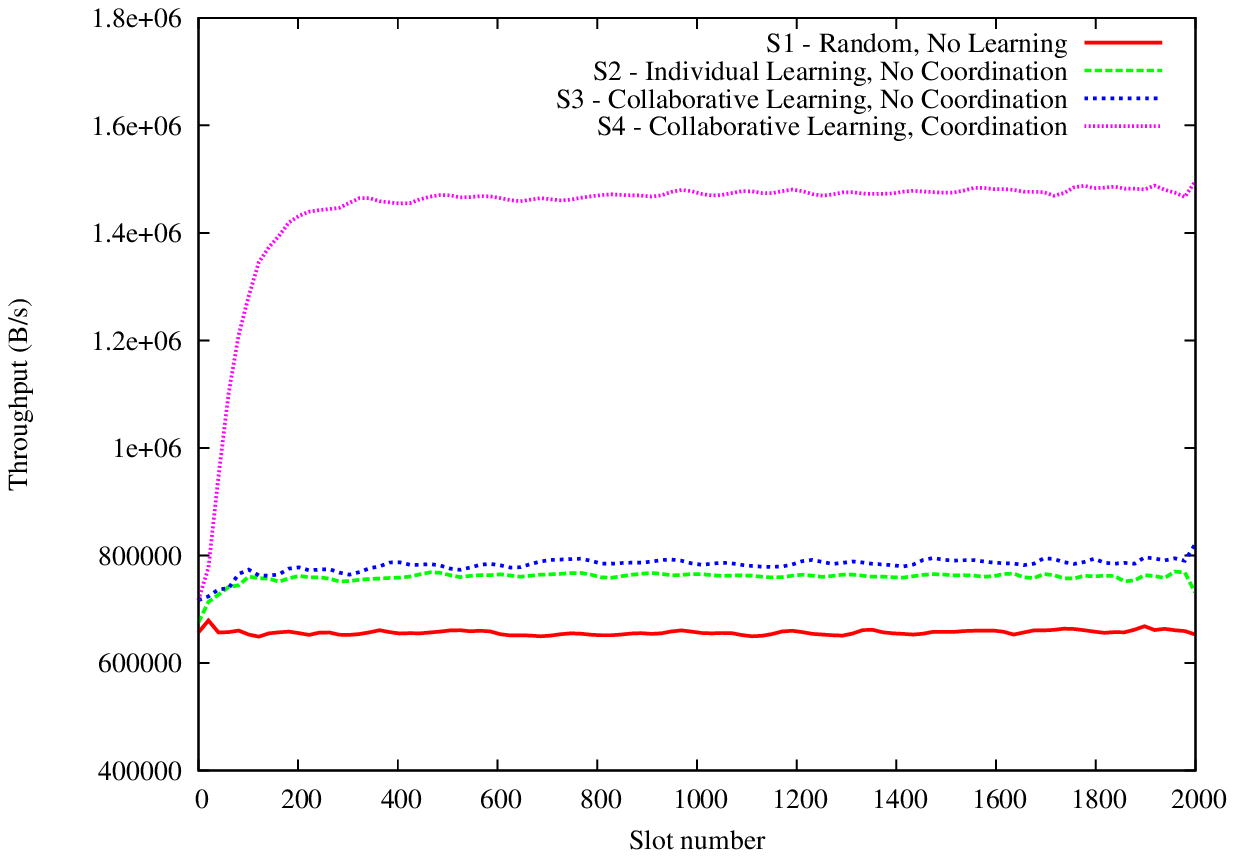}}
\subfigure[\label{thr2}] {\includegraphics[height = 7cm,
width=8.5cm]{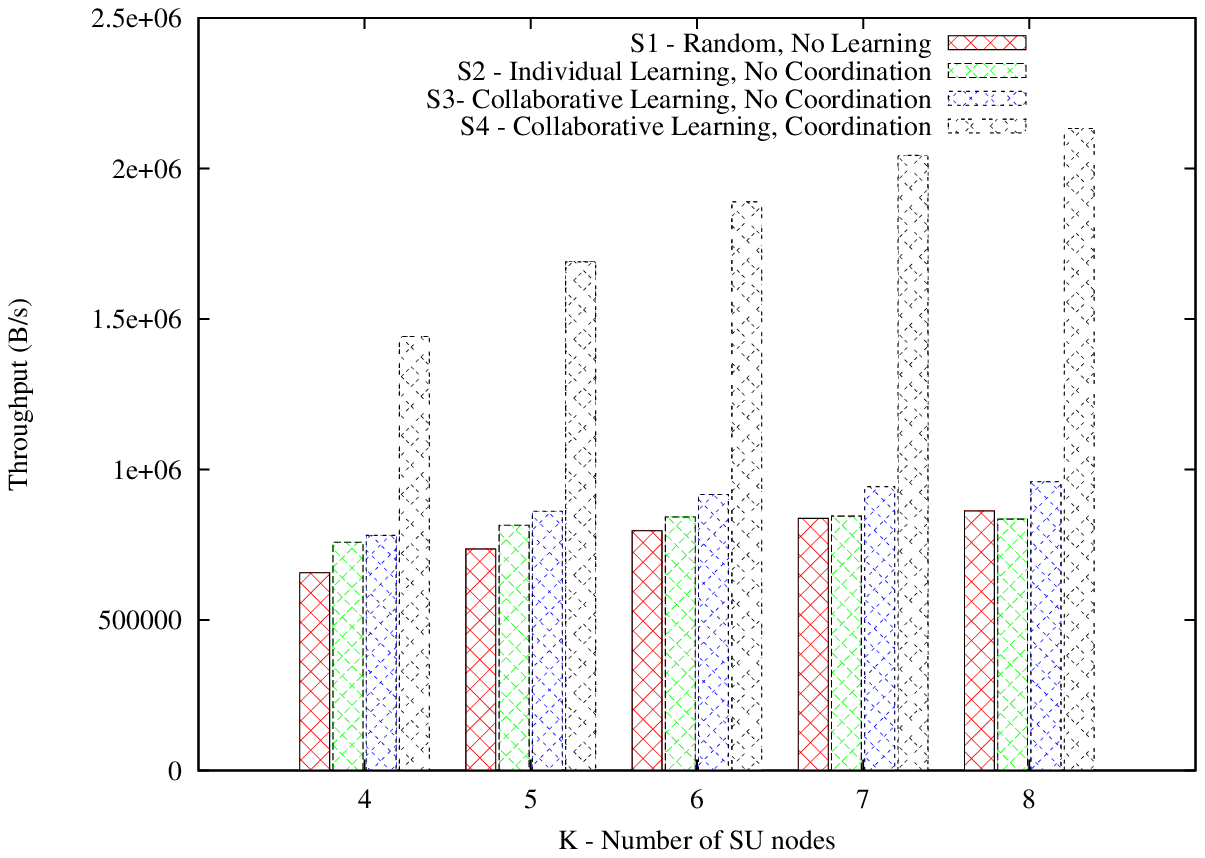}}
}
 \caption{The network throughput over simulation time in a scenario with $K$=4 is shown in Figure~\ref{thr1}. The network throughput as a function of the number of SUs (i.e. $K$) is shown in Figure~\ref{thr2}.  }
\end{figure*}

 Figure \ref{thr1} shows the network throughput as a function of the time slot $t$, averaged over 1000 simulation runs. As expected, the S1 scheme experiences the lowest throughput, since it does not take into account any mechanism to prevent SU and PU interference. On the other hand, both S2 and S3 schemes employ learning mechanisms to derive the PU occupation patterns of each channel, and thus are able to mitigate the interference caused by incumbent PU transmissions. Moreover, Figure \ref{thr1} shows that  the S3 scheme slightly enhances the S2 scheme since the usage of collaborative mechanism with reward sharing reduces the occurrence of wrong channel selection events due to local sensing errors. However, both S2 and S3 schemes do not include coordination mechanisms, and thus suffer of packet losses caused by SU interference i.e. by the fact that multiple SU transmitters are allocated on the same channel. The S4 scheme nullifies the harmful interference among SUs through Round Robin coordination, and thus provides the highest performance. Figure \ref{thr2} shows the average network throughput as a function of the number of $SU$ transmitters in the network. Again, Figure \ref{thr2} shows that the joint cooperative learning and cooperative scheme provides the highest performance over all the scenarios considered. 

%---------------------------------------Conclusion-----------------------------------------------
\section{conclusion}
\label{sec:concl}
In this paper, we have addressed the problem of Opportunistic Spectrum Access (OSA) in coordinated secondary networks. We have formulated the problem as a cooperative learning task where SUs can share their information about spectrum availability.  We have analyzed the case of symmetric secondary networks, and we have provided some fundamental
results on the performance of cooperative learning schemes. Moreover, we have proposed a general coordination mechanism based on the Hungarian algorithm to address the general case (i.e. both symmetric and asymmetric networks). We are planning to validate our approach on cooperative learning schemes through further theoretical analysis and Cognitive Radio testbed-based implementations.
%------------------------------------------------------------------------------------------------
%%%---------------------------------------Acknowledgment-----------------------------------------------
%\section*{Acknowledgment}
%%%----------------------------------------------------------------------------------------------------
%The authors would like to thank Damien Ernest
 
\bibliographystyle{unsrt}
\bibliography{cognitive}

%---------------------------------------Acknowledgement------------------------------------------
\section*{Appendix: Proofs}
%------------------------------------------------------------------------------------------------

We introduce and prove in this section technical results used  to justify the important results stated in this paper.  
\begin{lemme}[Regret, general upper bound]
\label{lemme_regret1}
{Let us consider $K\geq1$ Symmetric Secondary Users and $N\geq K$ Primary channels. The SUs are assumed to have limited observation abilities defined by their  parameters $\{\epsilon_n,\delta_n\}$ for every channel $n$. Assuming that the Secondary Network follows the Coordination Policy \ref{coord_1} to select and access the primary channels, relying on $UCB_1$ algorithm with parameter $\alpha>1$, then every SU suffers, after $t$ slots, an expected cumulated regret $R^{(k)}_t$ upper bounded such that:}
\begin{align}
R^{(k)}_t \leq \sum_{n\notin \mathcal{D}^{*}}\frac{\left(\bar{\lambda}^{*}-\lambda_{n}\right)\mathbb{E}\left[T_n\left(\left\lfloor \frac{t}{K}\right\rfloor+K-1\right)\right]}{K}
\end{align}
\noindent where $\mathbb{E}\left[T_n(t)\right]$ refers to the expected number of pulls of a given channel $n$ (by all SUs), and where the following notations were introduced: 
$$
\left\{
	\begin{array}{ll}  
			\lambda_n=\left(1-\epsilon_n\right)\mu_{n}\\
			\bar{\lambda}^{*}=\frac{\sum_{n\in\mathcal{D}^{*}}\lambda_n}{K}\\ 
  \end{array}
\right.
$$
\end{lemme}
\begin{proof}
We can upper bound the regret of a user $k$ as defined in Equation \ref{eq:regret} by the regret that he suffers at the end of the considered round of $K$ plays, i.e., 
$$
R^{(k)}_t \leq R^{(k)}_{\left\lfloor t/K\right\rfloor K+K-1}\leq\sum^{\left\lfloor t/K\right\rfloor}_{m=0}\sum^{K-1}_{p=0}\left(\bar{\lambda}^{*}-\mathbb{E}\left[r^{(k)}_{Km+p}\right]\right)
$$
\noindent where the sum $\sum^{K-1}_{p=0}\left(\bar{\lambda}^{*}-\mathbb{E}\left[r^{(k)}_{Km+p}\right]\right)$ which refers to the cumulated loss during the round of $K$ plays indexed by the round number $m$, can also be written as:
$$
\sum^{K-1}_{p=0}\left(\bar{\lambda}^{*}-\mathbb{E}\left[r^{(k)}_{Km+p}\right]\right)=\sum_{{n\in \mathcal{D}^{*}}}\lambda_n-\sum^{K-1}_{p=0}\mathbb{E}\left[r^{(k)}_{Km+p}\right]
$$
\noindent which justifies the second inequality. Notice that this sum is positive if and only if at least one  sub-optimal channel, $n\notin \mathcal{D}^{*}$, is selected among the best $K$ channels to be played during the round $m$.

Thus we can further upper bound the regret as follows:
$$
R^{(k)}_t\leq\sum^{\left\lfloor t/K\right\rfloor}_{m=0}\sum_{n\notin \mathcal{D}^{*}}\left(\bar{\lambda}^{*}-\lambda_n\right)\mathbb{P}\left(n\in \mathcal{A}^{(k)}_{Km}\right)
$$
\noindent where $\mathcal{A}^{(k)}_{Km}$ refers to the $K$ channels with the highest indexes evaluated at the round number $m$ evaluated by the $k^{th}$ SU. An inversion of the two sum leads to the following expression inequality: 
\begin{align}
\label{eq:lem1}
R^{(k)}_t\leq\sum_{n\notin \mathcal{D}^{*}}\left(\bar{\lambda}^{*}-\lambda_n\right)\sum^{\left\lfloor t/K\right\rfloor}_{m=0}\mathbb{P}\left(n\in \mathcal{A}^{(k)}_{Km}\right)
\end{align}
Finally, we notice that the three following equalities are verified:
%\mathbb{E}\left[T^{(k)}_n(t)\right]=
$$
\left\{
	\begin{array}{ll}  
			\sum^{\left\lfloor t/K\right\rfloor}_{m=0}\mathbb{P}\left(n\in \mathcal{A}^{(k)}_{Km}\right)=\mathbb{E}\left[\sum^{\left\lfloor t/K\right\rfloor}_{m=0}\textbf{1}_{\left\{n\in \mathcal{A}^{(k)}_{Km}\right\}}\right]\\
			\sum^{\left\lfloor t/K\right\rfloor}_{m=0}\textbf{1}_{\left\{n\in \mathcal{A}^{(k)}_m\right\}}=T^{(k)}_n\left(\left\lfloor t/K\right\rfloor K+ K-1\right)\\
			T^{(k)}_n\left(\left\lfloor t/K\right\rfloor K+ K-1\right)={T_n\left(\left\lfloor t/K\right\rfloor K+ K-1\right)}/{K}
  \end{array}
\right.
$$
\noindent where the second equality can be read as: the number of time a channel $n$ is selected by a user $k$, until the slot number $\left\lfloor t/K\right\rfloor K+ K-1$, is equal to the number of rounds the event ${\left\{n\in \mathcal{A}^{(k)}_{Km}\right\}}$ is verified. The third equality on the other hand, reminds us that in the context of symmetric users, all SUs share the same information vector and obtain the same index values. Consequently, if a channel $n$ is selected at a given round, it is played exactly once by every SU. In other words, the channel is selected $K$ times during a round of $K$ plays.
\par Thus substituting and combining the three previous equalities with Equation \ref{eq:lem1} leads to the stated result and ends this proof. 
\end{proof}
\begin{lemme}
\label{lemme_sample}
{Let us consider $K\geq1$ Symmetric Secondary Users and $N\geq K$ Primary channels. The SUs are assumed to have limited observation abilities defined by their  parameters $\{\epsilon_n,\delta_n\}$ for every channel $n$. Assuming that the Secondary Network follows the Coordination Policy \ref{coord_1} to select and access the primary channels, relying on $UCB_1$ algorithm with parameter $\alpha>1$, then every suboptimal channel $n$, after $t$ slots, has an expected number of pulls upper bounded by a logarithmic function of the number of iterations that:}
\begin{align}
\mathbb{E}\left[T_{n}\left(\left\lfloor \frac{t}{K}\right\rfloor K+K-1\right)\right]\leq \frac{4\alpha}{\Delta^2_n}\ln\left(t+K-1\right) + o\left(\ln(t)\right)
\end{align}

\end{lemme}
\begin{proof}
We start by a first coarse upper bound verified for all $u_n \in \mathbb{N}$: since every channel is to be sensed at least $K$ times, we can write:
\begin{align}
\begin{array}{ll} 
\mathbb{E}\left[T_{n}\left(\left\lfloor \frac{t}{K}\right\rfloor K+K-1\right)\right]\leq K+u_n \\ +K\sum^{\left\lfloor \frac{t}{K}\right\rfloor}_{m=u_{n}+1}\mathbb{P}\left(n\in \mathcal{A}_{Km} ; T_{n}(Km)>u_{n}+1 \right)
\end{array}
\end{align}

Since we have the following event inclusion: $$\left\{n\in \mathcal{A}_{mK}\right\}\subseteq \left\{B_{T_n}(m)\geq\min_{n\in\mathcal{D}^*}\left\{B_{T_n}(m)\right\}\right\}$$
We can write:
\begin{align}
\label{eq:ineg1}
\begin{array}{ll} 
\mathbb{E}\left[T_{n}\left(\left\lfloor \frac{t}{K}\right\rfloor K+K-1\right)\right]\leq  K+u_n \\ +K\sum^{\left\lfloor \frac{t}{K}\right\rfloor}_{m=u_{n}+K}\mathbb{P}\left(B_{T_n}(m)\geq\min_{n\in\mathcal{D}^*}\left\{B_{T_n}(m)\right\}\right)
\end{array}
\end{align}

In this last inequality, the joint event $\{T_{n}(Km)>u_{n}+1\}$ is left implicit to ease the notations. This will be the case in the next assertion. Moreover notice that: $\forall n\in\mathcal{D}^*, \ K \leq T_{n}(Km)\leq m $.
Since for all $\tau \in \mathbb{R^+}$ we have the following event inclusion: 
\begin{align}
\begin{array}{ll}
\left\{B_{T_n}(Km)\geq\min_{n\in\mathcal{D}^*}\left\{B_{T_n}(Km)\right\}\right\} \\ \subseteq \left\{B_{T_n}(Km)\geq \tau \right\} \cup \left\{\min_{n\in\mathcal{D}^*}\{B_{T_n}(Km)\}< \tau \right\}
\end{array}
\end{align}
We can write:
\begin{align}
\begin{array}{ll} 
\mathbb{E}\left[T_{n}\left(\left\lfloor \frac{t}{K}\right\rfloor K+K-1\right)\right]\leq  K+u_n \\ +K\sum^{\left\lfloor \frac{t}{K}\right\rfloor}_{m=u_{n}+K} \mathbb{P}\left(B_{T_n}(Km)\geq \tau \right) \\ 
+K\sum^{\left\lfloor \frac{t}{K}\right\rfloor}_{m=u_{n}+K}\mathbb{P}\left(\min_{n\in\mathcal{D}^*}\left\{B_{T_n}(Km)\right\}<\tau \right)
\end{array}
\end{align}

For the rest of the proof we assume that:
\begin{align}
\left\{
\begin{array}{ll}
u_n=u_n(t)=\frac{4\alpha\ln\left(\left\lfloor\frac{t}{K}\right\rfloor K+K-1\right)}{\Delta^2_n} \\ \tau=\min_{n\in\mathcal{D}^*}\left\{\lambda_n\right\}
\end{array}
\right.
\end{align}
then we prove that:
\begin{align}
\left\{
\begin{array}{ll}
\sum^{\left\lfloor \frac{t}{K}\right\rfloor}_{m=u_{n}+K} \mathbb{P}\left(B_{T_n}(Km)\geq \tau 				\right)=o\left(\ln\left(t\right)\right)\\
\sum^{\left\lfloor 		\frac{t}{K}\right\rfloor}_{m=u_{n}+K}\mathbb{P}\left(\min_{n\in\mathcal{D}^*}\left\{B_{T_n}(Km)\right\}<\tau 	\right)=o\left(\ln\left(t\right)\right)
\end{array}
\right.
\end{align}
%-------------------------------------------------------Step2
\par First, we start by the following term: $\mathbb{P}\left(B_{T_n}(Km)\geq \tau\right)$. Notice that if the event (including its implicit event)
$\left\{B_{T_n}(Km)\geq \tau ; T_n(Km)>u_n+1\right\}$ is verified then there exists an integer $s: \ u_n+1\leq s \leq m$ such that the real value verifies $B_{s}(Km)\geq \tau$. Consequently, we can write:
\begin{align}
\mathbb{P}\left(B_{T_n}(Km)\geq \tau ; T_n>u_n+1\right)\leq \sum^{m}_{s=u_n+1}\mathbb{P}\left(B_{s}(Km)\geq \tau\right)
\end{align}
Considering an index value computed as detailed in Equations \ref{eq:index1} and \ref{eq:index2}, we can write:
\begin{align}
\mathbb{P}\left(B_{s}(Km) \geq \tau\right)&= \mathbb{P}\left(\bar{W}_{s}(Km) \geq \tau-A_{s}(Km)\right) \notag \\
&=\mathbb{P}\left(\bar{W}_{s}(Km) - \lambda_n\geq \tau-\lambda_n-A_{s}(Km)\right)
\end{align}
Since $s>u_n+1$, then:
$$
\tau-\lambda_n-A_{s}(Km)\geq\Delta_n-\sqrt{\frac{\alpha\ln(Km)}{u_n}}\geq\frac{\Delta_n}{2}
$$
Consequently, we can write:
\begin{align}
\mathbb{P}\left(B_{s}(Km) \geq \tau\right)  &\leq \mathbb{P}\left(W_{s}(Km) - \lambda_n\geq \frac{\Delta_n}{2} \right) \\
&\leq e^{-2(\frac{\Delta^2_{n}}{4})s} \\
&\leq e^{-2\alpha\ln\left(mK+K-1\right)}\leq\frac{1}{(mK)^{2\alpha}}
\end{align}
\noindent where the second inequality is a concentration inequality known as Hoeffding's inequality \cite{Hoeffding1963}. The third inequality is once again due to the inequality $s>u_n+1$.
Finally assuming that $\alpha>1$,
\begin{align}
\sum^{\left\lfloor \frac{t}{K}\right\rfloor}_{m=u_{n}+K} \mathbb{P}\left(B_{T_n}(Km)\geq \tau \right) &\leq \sum^{\left\lfloor \frac{t}{K}\right\rfloor}_{m=u_{n}+K}\sum^{m}_{s=u_{n}+1}\frac{1}{(mK)^{2\alpha}}\\
&\leq \sum^{\infty}_{m=u_{n}+K}\frac{1}{(m)^{2\alpha-1}(K)^{2\alpha}}\\
&=C_{n,\alpha}=o\left(\ln(t)\right)
\label{eq:ineg2}
\end{align}
where $C_{k,\alpha}$ exist for $\alpha>1$, is finite and is defined as the limit of Reimann's serie: 
$\sum^{\infty}_{m=(u_{n}+K)}\frac{1}{(m)^{2\alpha-1}(K)^{2\alpha}}$
\\
%-------------------------------------------------------Step3

\par We deal know with the following term: $\mathbb{P}\left(\min_{n\in\mathcal{D}^*}\left\{B_{T_n}(Km)\right\}<\tau \right)$ (including the implicit event). In order to avoid confusing optimal channels and sub-optimal channels, for the rest of this proof, we denote by $n^{*}$ a channel that belongs to the optimal set $\mathcal{D}^*$. As for the previous proof, and since for any $\{T_{n^*}\}_{\min_{{n^*}\in\mathcal{D}^*}}$, $K\leq T_{n^*}\leq m$, if the event $\{\min_{{n^*}\in\mathcal{D}^*}\left\{B_{T_{n^*}}(Km)\right\}<\tau;K\leq T_{n^*}\leq m\}$ is verified then there exists a channel ${{n^*}\in\mathcal{D}^*}$ and an integer $s_{n^*}: \ K \leq s_{n^*} \leq m$ such that the real value verifies $B_{s_{n^*}}(Km)\leq \tau$. To ease notations we introduce $\mathbb{P}_{n^{*}}$ the considered event:
$$
\mathbb{P}_{n^*}=\mathbb{P}\left(\min_{n\in\mathcal{D}^*}\left\{B_{T_{n^*}}(Km)\right\}<\tau ; T_{n^*}>K\right)
$$
Consequently we can write:
\begin{align}
\mathbb{P}_{n^*} \leq \sum_{{n^*}\in\mathcal{D}^*}\sum^{m}_{s_n=K+1}\mathbb{P}\left(B_{s_{n^*}}(Km)< \tau\right)
\end{align}
Notice that for any ${{n^*}\in\mathcal{D}^*}$: 
$$
\min_{{n^*}\in\mathcal{D}^*}\left\{\lambda_{n^*}\right\}-\lambda_{n^*}\leq0
$$
Consequently, as for the previous proof, relying on Hoeffding's inequality, we can write: 
\begin{align}
\mathbb{P}_{n^*} &\leq \sum_{n^{*}\in\mathcal{D}^*}\sum^{m}_{s_{n^*}=K+1}\mathbb{P}\left(\bar{W}_{s_{n^*}}(m)-\lambda_{n^*}<-A_{n^*} \right)\\
&\leq \sum^{m}_{s_{n^*}=K+1}K e^{-2A^{2}_{n^{*}} s_{n^{*}}}\\
&\leq \sum^{m}_{s_{n^*}=K+1}K e^{-2\alpha\ln\left(Km\right)}\\
&\leq \frac{1}{(Km)^{2\alpha-1}}
\end{align}
Finally, we can write:
\begin{align}
\sum^{\left\lfloor \frac{t}{K}\right\rfloor}_{m=u_n+1}\mathbb{P}_{n^*} &\leq \sum^{\left\lfloor \frac{t}{K}\right\rfloor}_{m=u_n+1}\frac{1}{(Km)^{2\alpha-1}}\\
&\leq \sum^{\infty}_{m=u_n+1}\frac{1}{(Km)^{2\alpha-1}}\\
&=C_{{n^*},\alpha}=o\left(\ln(t)\right)
\label{eq:ineg3}
\end{align}
\noindent where $C_{{n^*},\alpha}$ exist for $\alpha>1$, is finite and is defined as the limit of Reimann's serie:
$
\sum^{\infty}_{m=u_n+1}\frac{1}{(Km)^{2\alpha-1}}
$.
\par Finally, since: $\left\lfloor \frac{t}{K}\right\rfloor K+K-1\leq t+K-1$, combining Inequalities \ref{eq:ineg1}, \ref{eq:ineg2} and \ref{eq:ineg3}, we can finally write:
\begin{align}
\begin{array}{ll} 
\mathbb{E}\left[T_{n}\left(\left\lfloor \frac{t}{K}\right\rfloor K+K-1\right)\right]\leq \frac{4\alpha}{\Delta^2_n}\ln\left(t+K-1\right)+o\left(\ln(t)\right)
\end{array}
\end{align}
Which ends the proof.
\end{proof}

%\end{thebibliography}

\end{document}